\documentclass[sigconf, screen]{acmart}

\AtBeginDocument{%
  \providecommand\BibTeX{{%
    \normalfont B\kern-0.5em{\scshape i\kern-0.25em b}\kern-0.8em\TeX}}}

\begin{document}

\title{The Ifs and Buts of the Development Approaches for IoT Applications}
\author{Saitel Daniela Agudelo-Sanabria}
\email{saitel.agudelo@tum.de}
\orcid{0002-8287-5198}
\affiliation{%
  \institution{Technical University of Munich}
    \country{Germany}
}

\author{Anshul Jindal}
\email{anshul.jindal@tum.de}
\orcid{0002-7773-5342}
\affiliation{%
  \institution{Technical University of Munich}
    \country{Germany}
}

\renewcommand{\shortauthors}{Agudelo et al.}

\begin{abstract}
  The recent growth of the Internet of Things (IoT) devices has lead to the rise of various complex applications where these applications involve interactions among large numbers of heterogeneous devices. An important challenge that needs to be addressed is to facilitate the agile development of IoT applications with minimal effort by the various parties involved in the process. However, IoT application development is challenging due to the wide variety of hardware and software technologies that interact in an IoT system. Moreover,  it involves dealing with issues that are attributed to different software life-cycle phases: development, deployment, and progression. 
  
  In this paper, we examine three IoT application development approaches: Mashup-based development, Model-based development, and Function-as-a-Service based development. The advantages and disadvantages of each approach are discussed from different perspectives, including reliability, deployment expeditiousness, ease of use, and targeted audience. Finally, we propose a simple solution where these techniques are combined to deliver reliable applications while reducing costs and time to release.
\end{abstract}



\maketitle
\section{Introduction}
\label{sec_introduction}
The Internet of Things (IoT) is a network that enables \textit{things} to monitor their surroundings, interact with humans and other \textit{things}, and carry out all sorts of tasks with little to none human intervention~\cite{Mala2019}. Sensors, controllers, actuators, and any other device capable of establishing a connection to the Internet are called a \textit{thing}~\cite{Waher1386}.

The application areas of the Internet of Things are as heterogeneous as the devices and applications that compound it. Industries, in general, can take advantage of IoT to predict device failure, manage their supply chain, and optimize their manufacturing and administrative processes~\cite{Casado-Vara2019, Belli2019, ArvindRavulavaru2018}. The retail industry is using IoT to enable new customer experiences. Some examples are Amazon go smart store~\cite{Amazon2016}, and Alibaba's smart warehouse~\cite{BusinessInsider2017}. The agriculture industry benefits from the possibility of automating the crop yield measuring, livestock monitoring, soil quality control, irrigation, and other production processes~\cite{Pattnaik2019}. This kind of services are offered by companies like~\cite{ KaaIoTTechnologiesLLC} and~\cite{SoftwebSolutionsInc}. More applications examples can be found in automotive~\cite{Syafrudin2018}, nuclear~\cite{Susila2018}, aerospace~\cite{Correia2019}, and military~\cite{Lan2012} industries, among others.
Furthermore, we find IoT applications in government, heath, and security systems. Local governments' challenges such as pollution control, waste handling, energy management, disaster prevention and response, parking assistance, and traffic re-routing, find solutions in the context of IoT enabled smart cities~\cite{Krylovskiy2015}. Undoubtedly, IoT plays an every-time more important role in our daily lives. 




The existence of a diversity of scenarios where IoT is a leading actor is only possible with the cooperation of multiple technologies and protocols. The complexity of the architecture and the variety of technologies combined in an IoT system is the most visible challenge of IoT application development. Furthermore, developers must deliver applications that behave consistently under different operating conditions, that might change at runtime~\cite{Mala2019}. In some cases, machine-human interfaces are not necessary or available, so we need to enable other communication channels in case of failure~\cite{Mala2019}. At the current time, there exists many different development strategies which are grouped under two categories: Model-based and mashup development. However, there is no consensus on a general development strategy. On the one hand, software engineering, as a discipline, provides the structure and the tools necessary to describe every aspect of an IoT system~\cite{Mala2019}. However, it lacks the expediteness and ease of use of mashups. On the other hand, mashups support for black-box development, in which the programmer does not need to develop the components of the system from scratch nor know their implementation in detail~\cite{Ogrinz2009}. Only the inputs and outputs need to be identified to include the element in the interconnected network of components that make up the application.

Furthermore, with the introduction of the concept of serverless since the launch of AWS Lambda in 2014~\cite{Alex_Handy_2014}, it has gained higher popularity and more adoption in different fields. Function-as-a-Service (FaaS) is a key enabler of serverless computing~\cite{wg2018cncf}. In FaaS, an application is decomposed into simple, standalone functions that are uploaded to a FaaS platform for execution. FaaS offers a new way for the development of the IoT applications where one can imagine all the IoT devices to be part of a FaaS platform and then functions are scheduled on each of the devices depending on their computational capabilities and requirements. 

Our key contributions are: 
\begin{itemize}

\item  We provide an overview of the current trends in programming models and tools used to build IoT applications.

\item We introduce the extension of development of IoT applications based on the FaaS approach using Google Cloud Functions (on Google cloud).

\item We examine three IoT application development approaches: Mashup-based development, Model-based development, and Function-as-a-Service based development using a real world example from different perspectives, including reliability, deployment expeditiousness, ease of use, and targeted audience. 

\item We propose a simple solution where the development techniques are combined to deliver reliable applications while reducing costs and time to release.
\end{itemize}

The rest of the paper is organized as follows. In Section~\ref{sec_development_approaches}, the different IoT application development approaches are described along with the IoT system layered architecture. A practical IoT application example is presented using all the approaches in   section~\ref{sub_appExample}. Section~\ref{sec_discussion} compares the methodologies and discusses the ways in which they can be combined to allow for transparent, simple, fast, and robust development. Lastly, section~\ref{sec_conclusion} concludes the paper.

\section{IoT application development Approaches}
\label{sec_development_approaches}

\begin{figure}[ht]
\centerline{
\includegraphics[width=\columnwidth]{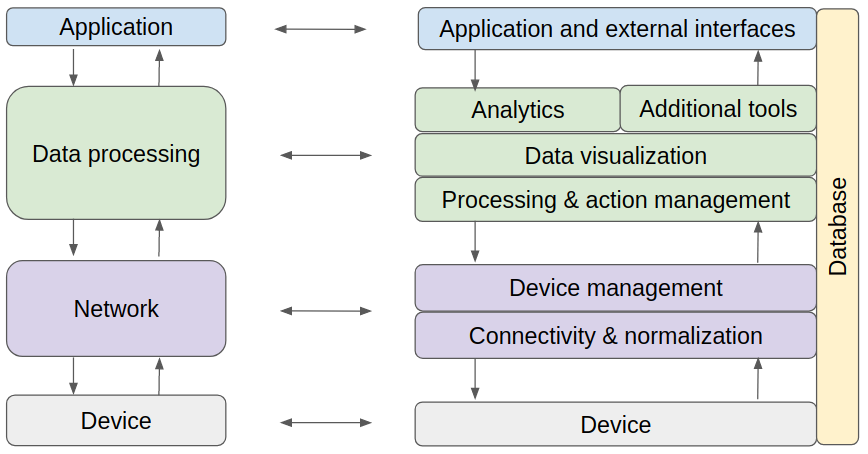}
}
\caption{A general layered IoT system architecture. Devices in the \textit{Device Layer} gather data and pass it to the \textit{Network Layer}, where it is transmitted to the upper layer. The data is then stored and processed in the \textit{Data Processing Layer}, passed to the \textit{Application Layer} that enables machine-to-human communication through front-end applications and machine-to-machine communication through appropriate communication protocols~\cite{Mala2019,Scully2016}.}

\label{fig_IoTLayers}
\end{figure}

An IoT system can be described using four layers that group the main processes: the Device, Network, Data Processing, and Application~\cite{Mala2019}. A general IoT system layered architecture is depicted in Figure~\ref{fig_IoTLayers}. Every IoT application requires things: physical equipment such as sensors, actuators, and controllers, which are grouped in the \textit{Device Layer}. For example, sensors measuring the heart rate, oxygenation, and temperature of a patient. \textit{Network Layer} is responsible for transmitting the monitored data to the cloud. In the \textit{Data Processing Layer} the data is transformed into information that can be used to make decisions. For example, if the heart rate of the patient under observation shows an abnormal raise, the IoT system can notify the doctors. The \textit{Application Layer} act as a interface for a user to understand or convey the information. For example, sending a warning message or directly by making a call to the doctor. The decision from the doctor or the user is transmitted back down through the hierarchy and is executed by the devices in the \textit{Device layer}.


In the following subsections, we introduce three IoT application development approaches.

\subsection{Mashup-based development}
\label{sec_mashup}

Mashups applications are composites of existing assets that cooperate to deliver new content, with reduced time-to-production, complexity and cost. The development process begins when a new business opportunity is identified. The developers then look for all existing resources that can be integrated to construct the new application. It is important to note that mashups are not necessarily final products nor have always a user interface~\cite{Ogrinz2009}. The components are represented as black-boxes, with inputs and outputs specified in the APIs. The developers may need to build some functionality into a new block or perform adaptations of the selected APIs. Once the components are connected, the application can be tested and opened to the public. Then, new applications can be built on top of this one, and the development cycle can begin again. This process is described by Michael Ogrinz in~\cite{Ogrinz2009} as the circle of mashups.

\begin{figure}[ht]
\centerline{
\includegraphics[width=\columnwidth]{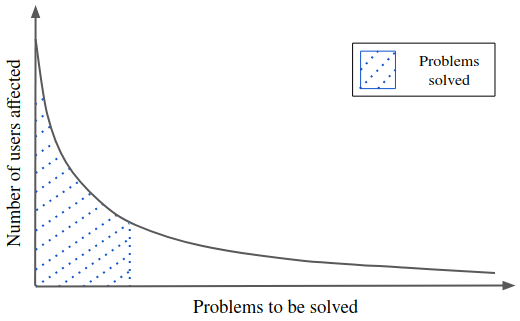}
}
\caption{\textbf{Long tail problem in software development}. Development teams focus on solving a small number of problems that affect the majority of users, while a large number of very specific problems remain unsolved. Reinterpretation of the figure in~\cite{Ogrinz2009}}
\label{fig_longTail}
\end{figure}

Development teams typically focus on solving a small number of problems that affect the majority of users, while a large number of very specific issues remain unsolved. This is known as \textit{The Long Tail Problem} (See Fig~\ref{fig_longTail}). Mashups were previously described as a solution to  this problem in the context of enterprise applications for internal use ~\cite{Ogrinz2009}. In the case explained in ~\cite{Ogrinz2009}, IT teams would develop the most important features, and allow non-IT teams to build mashup applications to address their specific requirements within a controlled environment where the functionality is presented as widgets. This asseveration can be extended to software development in general. Today, many successful service providers are exposing APIs, allowing for tailor-made mashup applications that improve user experience, open communication channels, and are ultimately translated in increasing revenue~\cite{Chow2007}.

The development of mashups can be manual or assisted by tools~\cite{Yu2008}. The manual process requires a significant level of knowledge of the technologies involved. That is because mashup applications integrate web-based artifacts such as RSS/Atom feeds and HTML data, and other types of resources like databases, binary and XML files~\cite{Ogrinz2009}, which implies that different applications might provide different formats to retrieve data. Therefore, the data will need to be parsed, so that it can be understood by all the parties associated. Additionally, if a user interface is to be provided, the developer needs to take care of its functionality. 

We focus our discussion on assisted development of IoT applications. A study conducted in 2010 pointed out that mashup tools were not sufficiently easy to handle from the end-programming perspective~\cite{Patel2010}. Today, mashup tools have evolved from the end-programming paradigm to offer different programming schemes and automation levels according to the targeted audience, which might include non-programmers, people with little experience in programming, and experienced developers~\cite{Aghaee2012}. The final product might be a non-executable design, a prototype with mock elements, or a deployable application depending on the tool~\cite{Aghaee2012}. A great characteristic of mashup tools that appeals to both programmers and non-programmer users is the community built around this tools, which is not only capable of offering guidance but also of extending the tool features. 

Typical functionalities supported by mashup tools are: 
\begin{itemize}

\item Creating, configuring, and connecting nodes in a user-friendly graphical editor.
\item Data gathering, combination, and transformation. 
\item Scripting/Coding.
\end{itemize}
IoT frameworks usually offer an extended set of functionalities, such as domain-specific language and coding aids to support the development process. Some examples of IoT frameworks are: \textit{Watson IoT Platform}~\cite{IBM2019}, Kaa~\cite{KaaIoTTechnologiesLLCa}, Crosser~\cite{Crosser}, Thingsboard~\cite{Thingboard2020}, and MindSphere~\cite{Siemens}. Some IoT mashup tools examples are Paraimpu~\cite{Paeaimpu2016}, Total.js Flow~\cite{Flow} and Node-RED~\cite{Node-RED}. Paraimpu is designed as a social platform. It offers a web interface where users can configure a predefined set of sensors and actuators, share and subscribe to \textit{things} shared by their contacts~\cite{Piras2014}. Total.js Flow and Node-RED are open-source visual programming tools that support IoT, web, and REST applications. In both, users can drag, drop, and connect nodes in a graphical editor. Total.js Flow shows a real-time representation of data traffic between connected nodes and node errors~\cite{Flow}. We present an application developed using Node-RED in subsection~\ref{sub_mashupExample}.

\subsection{Model-driven development}
\label{sec_model}

Software engineering allows for robust IoT application delivery, by providing developers with a structured development process~\cite{Mala2019}. The specification of a software structure and behaviour usually involves Unified Modelling Language (UML) diagrams. The structure of a system can be described using class diagrams, component diagrams, composite structure diagrams, object diagrams, package diagrams, deployment diagrams, and profile diagrams. To describe the behavior of a system one can use activity diagrams, state diagrams, communication diagrams, interaction overview diagrams, sequence diagrams, timing diagrams and use case diagrams. A complete description of each diagram can be found in~\cite{Rumbaugh2004}. These, sometimes overwhelming, amount of representations of the system creates a complete definition from different levels of abstraction.

When it comes to IoT, models extending UML can become as specific as the applications and technologies involved. Nastic et al., for instance, proposed PatTRICIA, a programming model based on \textit{Intents}, which represent tasks, and \textit{Intent Scopes}, which establish the entities in which the tasks are executed. Barbon et al. proposed ASIP for IoT development in Arduino platforms, introducing the concept of \textit{services} as devices that communicate with the board via textual messages~\cite{Barbon2016}. Nguyen, et al. proposed \textit{FRASAD}, a framework that employs sensor-nodes as the central notion~\cite{Nguyen2015}. 

The main disadvantage of this development approach is that it is time-consuming. Not only because of the many abstraction levels in which systems should be described but also because of the granularity of the structures that can be reused i.e classes and code snippets~\cite{Rumbaugh2004}. This is one of the points where developers can benefit from the integration of mashup development to their workflows (See Section~\ref{sec_discussion}). Nonetheless, model-based development tools offer another solution: Code generation. Code can be automatically generated from different kind of diagrams. Much research has been conducted to propose study the efficiency of the generation process using different kind of diagrams, such us sequence diagrams~\cite{Kundu2013}, activity and sequence diagrams~\cite{Viswanathan2016}, state chart diagrams~\cite{Sunitha2019}, and class diagrams~\cite{Sejans2012}. Wang et al. proposed a fog-based model for IoT applications in the field of smart grids~\cite{Wang2018}. 

There are many code generation tools for different programming languages. Some examples are Papyrus~\cite{EclipseFoundation} for C++ and Java, ThingML for C, C++, Java, and JavaScript~\cite{Harrand2016}, and Visual Paradigm for 17 languages including C\#, Java, Python, PERL, and Ruby~\cite{VisualParadigm}. Code generators create mappings between certain diagram structures and lines of code. The ability of these tools to translate one structure into another is variable; some of them only produce skeletal code, while others can match almost every structure.

\subsection{Function-as-a-Service based development}
\label{sec_faas_theory}
Function-as-a-Service (FaaS) provides an attractive cloud model since it facilitates application development and reduces application costs. Instead of developing application logic in the form of services and managing the required resources, the application developer implements fine-grained functions connected in an event-driven application and deploys them into the FaaS platform~\cite{wg2018cncf}. The platform is responsible for providing resources for function invocations and performs automatic scaling depending on the workload.  The functions can be closely integrated with other services, e.g., cloud databases, authentication and authorization services, and messaging services. These services are sometimes called Backend-as-a-Service (BaaS).  BaaS are the third-party services that replace a subset of functionality in a function and allow the users to only focus on the application logic~\cite{lane2015overview}. Since functions are stateless, the state of the application is stored in databases.
\begin{figure*}
\centerline{
\includegraphics[width=\textwidth]{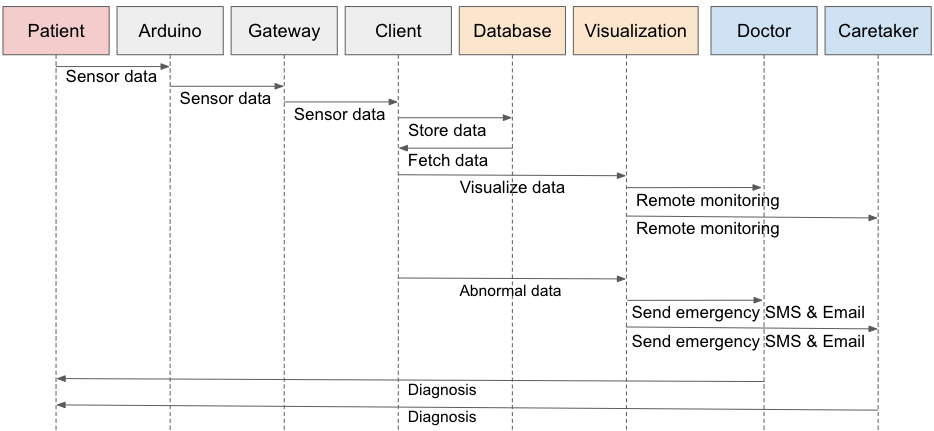}}
\caption{Sequence diagram for the example remote health monitoring system application}
\label{fig_seqDiagram}
\end{figure*}
Building a secure, scalable, performant and managed application for an IoT system seems like a huge challenge, however, there have been growing development to run FaaS based functions on the IoT devices. For example, 
in the IoT Greengrass system of Amazon~\cite{AWSIoTGr10:online}, it is possible to integrate end devices with cloud resources in an IoT platform and application Lambda functions are deployed to the end devices. However, this approach is limited to single applications on the edge and a static distribution of computation. The integration of IoT systems for general FaaS applications will require an extension of the FaaS platform across heterogeneous devices. This kind of development offers several advantages:

\begin{itemize}
\item  It reduces operational and system administration costs.
\item  It reduces the development and deployment costs, and provides faster time to market. 
\item  It is highly scalable and fault tolerant.
\end{itemize}
Therefore, FaaS provides an alternate easy for developing IoT applications.

\section{Application Overview}
\label{sub_appExample}
In this section we present an overview of an example application considered for demonstrating all the development approaches.

We consider an IoT-based health monitoring system that uses a sensor to gather the heart signal of a patient at a particular sample rate for example 100Hz. Fig.~\ref{fig_seqDiagram} shows the sequence diagram for the application. The sensor data is then published through an Ardunio device to a remote MQTT gateway broker to a particular topic. A client subscribing to that topic, store the received measurements in a NoSQL database such as MongoDB. Further, this data is  analyzed to obtain the following metrics: 
\begin{itemize}
    \item Beats per minute (BPM)
    \item Interbeat interval (IBI)
    \item Standard deviation of RR intervals (SDNN)
    \item Standard deviation of successive differences (SDSD)
    \item Root mean square of successive differences (RMSSD)
    \item Proportion of successive differences above 20ms (pNN20)
    \item Proportion of successive differences above 50ms (pNN50)
    \item Median absolute deviation of RR intervals (MAD)
\end{itemize}
The calculated metrics metrics are presented in the visualization using graphs for remote monitoring and the abnormalities are highlighted for the doctor and caretaker for quick actions. 
 This can be also visited at any time by the doctor and the caretaker.

In practice, a sensor would measure the change in the skin color as the blood is pumped. However, in this work, we emulated the heart signal sensing using the data in~\cite{Vangent2016}, which had a sample rate of 100 values per second. We utilize  Eclipse Mosquitto an open source MQTT broker on the Google Cloud over a virtual machine with 2 vCPUs and 7.5 GB memory. The NodeRED was also hosted on the Google Cloud over a virtual machine with 2 vCPUs and 7.5 GB memory. Python  \textit{pyHeart}~\cite{VanGent2018, VanGent2019} package is used for performing the analysis. For creating functions as part of FaaS development strategy, we used Google Cloud Functions on the Google cloud. We configured each function with a memory size of \texttt{128mb} and timeout of \texttt{60} seconds.


\begin{figure*}
\centerline{
\includegraphics[width=\textwidth]{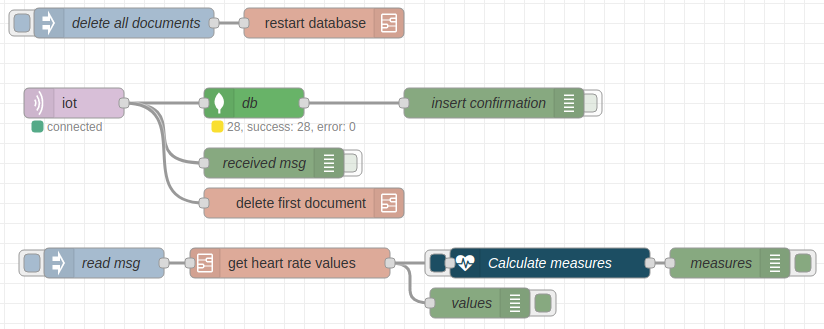}
}
\caption{\textbf{Flow of the proposed health monitoring system using mashup based development approach.}}
\label{fig_muExample}
\end{figure*}
\section{Methodology}
In this section, we present the detailed development design and architecture of the example application using the three development approaches.

\subsection{Mashup-based development}
\label{sub_mashupExample}

We implemented the application system proposed in Section~\ref{sub_appExample} using NodeRED, a mashup programming tool in which the behavior of an application can be described as a directed graph composed of wired nodes~\cite{Node-RED}. The nodes might be APIs, hardware devices, databases, online services, and other applications. A set of related nodes is called a flow. NodeRED provides an online editor, which can be accessed in a browser. It allows for effortless addition of nodes using drag and drop gestures. The nodes can be connected by wires that represent the flow of data. Every resource-node can be configured by filling up a form that requests for relevant data, such as credentials, web addresses, and ports. Once the node information is given, the application can be deployed to the Node.js runtime environment and executed~\cite{Node-RED}. 

Since the project was open-sourced back in 2013, and thanks to the participation of its evergrowing community, the tool has added multiple features and node libraries~\cite{Node-RED}. Today it has nodes for hardware integration, input/output handling, social media, storage, time, data generation, processing, analysis and parsing, among others. The tool can be run locally, in a docker, on a programmable board, or in the cloud, and the flows can be exported and shared as JSON files~\cite{Node-RED}. 

We configured the  built-in \textit{MQTT subscriber} node to receive the messages from the MQTT broker. We further stored the incoming messages to MongoDB for long-term availability using the \textit{db} node in NodeRED and performed a continuous analysis of the incoming data using the \textit{python-function} node where the python script for analyzing the data was run. The flow of the system is presented in Fig.~\ref{fig_muExample}.

The representation is composed by three flows. The flow that appears in the upper part of the image is used mainly for debug purposes. It deletes all the documents in the database. The flow in the middle of the image allows us to store a fixed number of entries in the database. It receives messages from a MQTT broker and stores them in the database. Once a user-defined threshold is met, the first stored value is removed, ensuring that the database keeps a manageable size. Lastly, the flow bottom at the bottom of the image, retrieves the measurements from the database and pass them as an argument to a python script that filters them and uses them to calculate the heart-related metrics presented in Section~\ref{sub_appExample}.


\subsection{Model-based development}


\begin{figure}[ht!]
\centerline{
\includegraphics[width=1\columnwidth]{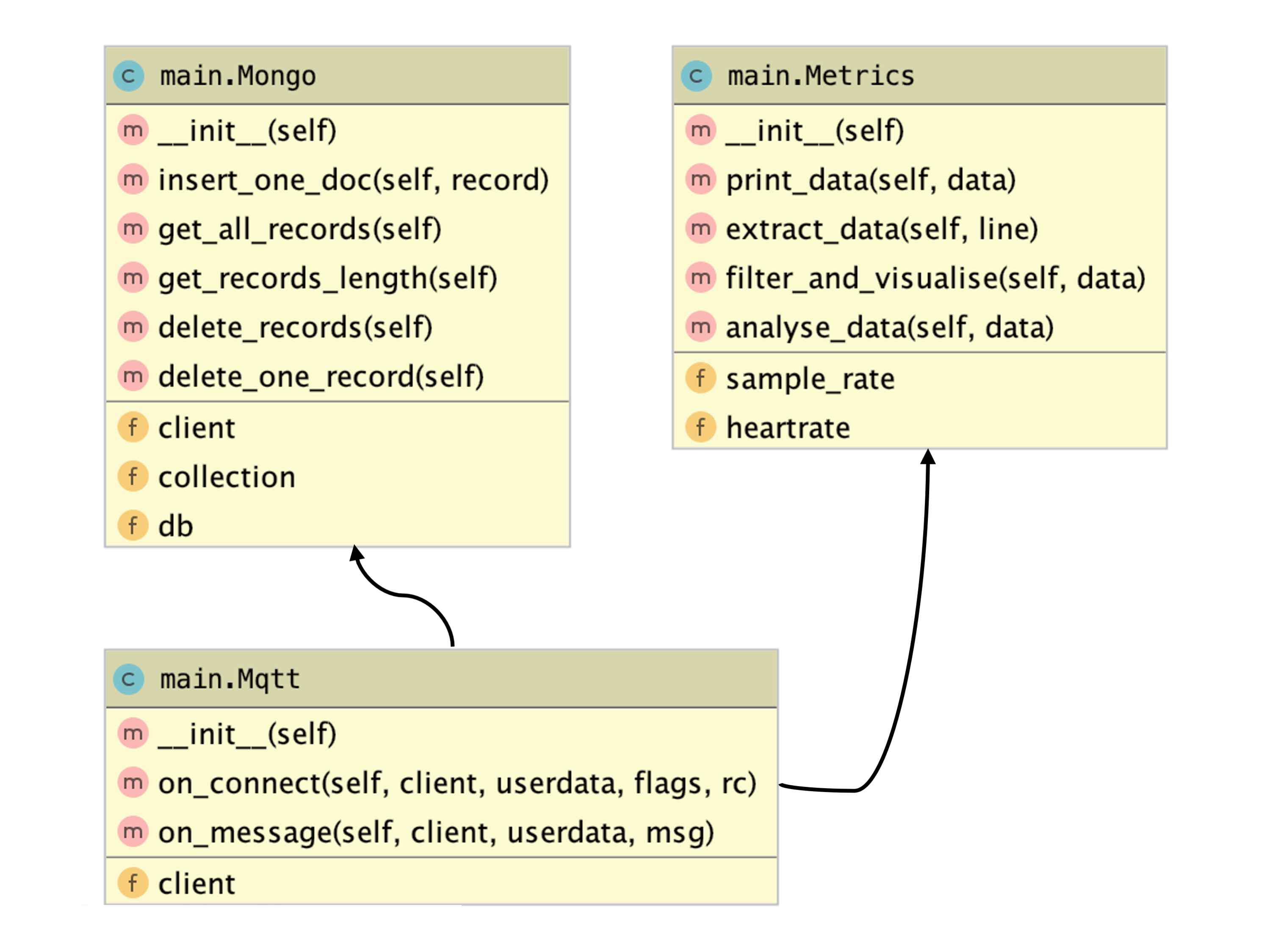}}
\caption{Class diagram for the sample remote health monitoring system application}
\label{fig_classDiagram}
\end{figure}
The class diagram of the proposed application system in Section~\ref{sub_appExample} is shown in Fig.~\ref{fig_classDiagram}. This diagram explains the abstraction of the system from a programming point of view. We observe the three different classes and their relationships.

\begin{itemize}
    \item Class Mongo: This class is responsible for interacting with Mongo database and performing different operations like: inserting the document, getting all the records, deleting the records. 
    
    \item Class Metrics: This class is responsible for reading the data, converting it into desired form and then analyzing it for calculating various measurements presented in Section~\ref{sub_appExample}.
    
    \item Class Mqtt: It is responsible for connecting to the remote MQTT broker and when a message is received then calling Mongo class object for storing it in the database and Metrics class object for  calculating various measurements.
    
\end{itemize}
\subsection{Function-as-a-Service based development}
\label{sec_faas}
\begin{figure}[ht!]
\centerline{
\includegraphics[width=1\columnwidth]{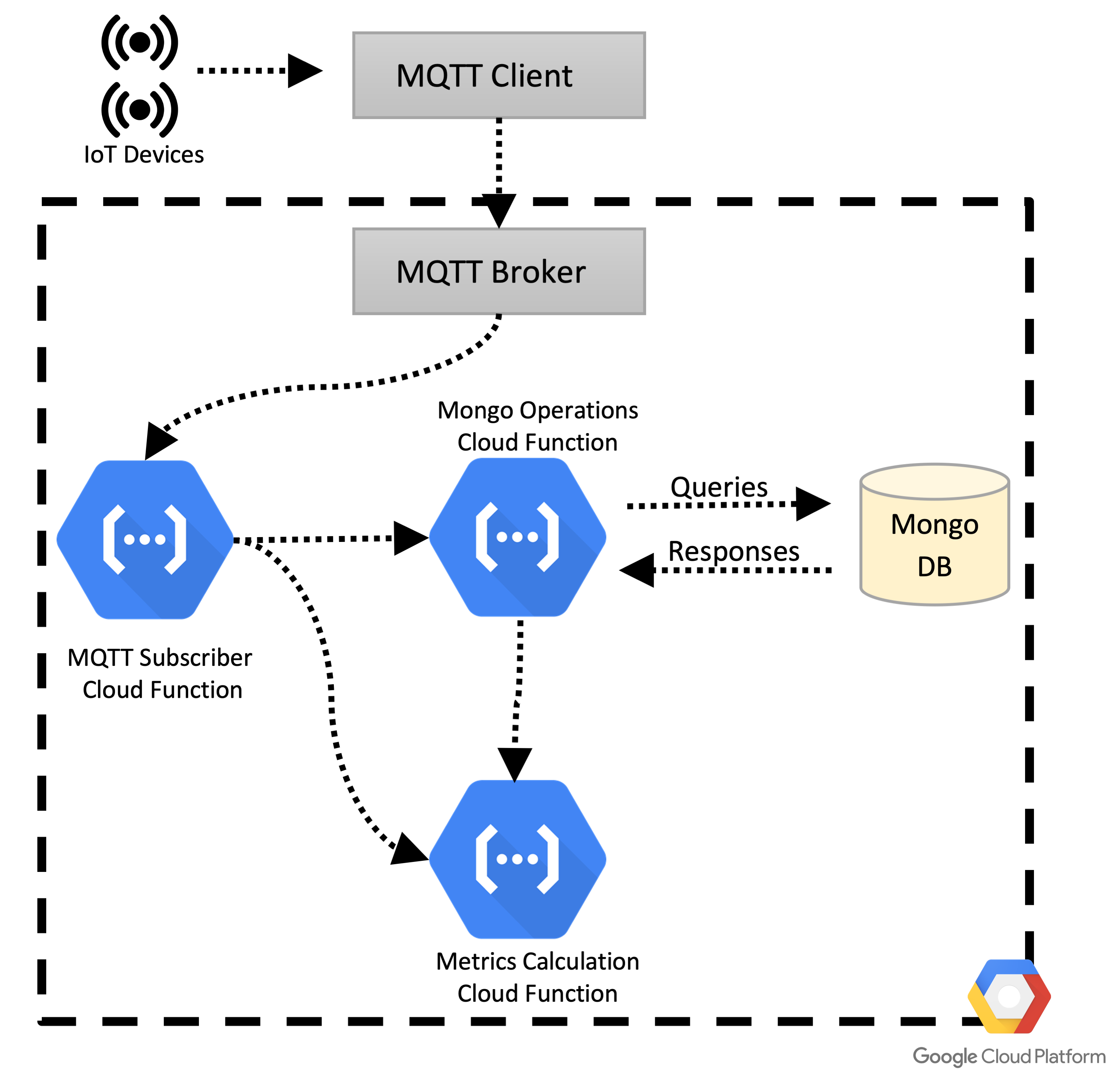}}
\caption{Function-as-a-Service based modeling of the proposed health monitoring system.}
\label{fig_faasmodel}
\end{figure}

 Figure~\ref{fig_faasmodel} shows the modeling of the proposed application system in Section~\ref{sub_appExample} using Function-as-a-Service on Google Cloud Platform. The system consist of three cloud functions:

\begin{itemize}
    \item Mongo Operations Cloud Function: This Google Cloud Function (GCF) is responsible for interacting with Mongo database and performing different operations like: inserting the document, getting all the records, deleting the records. 
    
    \item  Metrics Calculation Cloud Function: This GCF is responsible for reading the data, converting it into desired form and then analyzing it for calculating various measurements presented in Section~\ref{sub_appExample}.
    
    \item  MQTT Subscriber Cloud Function:: It is responsible for connecting to the remote MQTT broker and when a message is received then invoking \textit{Mongo Operations Cloud Function} for storing the data in the database and \textit{Metrics Calculation Cloud Function} for  calculating various measurements.
    
\end{itemize}

\section{Discussion}
\label{sec_discussion}

The lack of a systematic approach to develop IoT applications is pointed out as an issue~\cite{Mala2019, Prehofer2015}. At firsts, one might think this fact obeys to the absence of efforts to put up a modeling scheme. However, after careful review, one discovers that is the abundance of modeling approaches and not the lack, what generates the issue. In this section we compare mashup development with model-based development and discuss the possible integration on these two approaches. A summary is provided in~\ref{tab_comparison}.

\begin{table*}[hbt!]
\caption{Comparison of different development approaches at various attributes.}
\label{tab_comparison}
\begin{tabular}{|l|l|l|l|l|}
\hline
  & \textbf{Attribute}                                        & \textbf{Mashup-based}               & \textbf{Model-based}   & \textbf{FaaS-based} \\
  \hline
1 & End-users are able to produce applications       & Yes                     &        No &        No        \\
2 & Can reuse existing resources                     & At application level & At class level & Yes\\
3 & Platform independent                             & No                    &      Yes      & Yes    \\
4 & Suitable for critical systems                    & No*                 &      Yes      & Yes    \\
5 & Maintenance can be schedule in a predictable way & No*                    &     Yes      & Yes     \\
6 & Can warranty high availability                   & No*                    &   Yes & Yes \\    
    \bottomrule
\end{tabular}%
\par
\footnotesize{* Unless all resources are owned by the organization developing the application} 
\end{table*}

Let us start by comparing model-based and mashup diagrams. No structure diagram, except for the deployment diagram, can be compared to a mashup representation. These kind of diagrams describe the deployment of artifacts on nodes, which correspond to a location~\cite{Rumbaugh2004}. Nodes in mashup applications might indeed correspond to physical devices, but different nodes can also correspond to distinct applications on the same server. Similarly, in FaaS-based approach different functions can also correspond to physical devices or to distinct application functions. 

From the behavioral UML diagrams, activity diagrams are the closest system representation to mashups. Activities model behavior, and are connected by control or data flows~\cite{Rumbaugh2004}. Mashups nodes have a defined behavior and are connected by data flows. The latter was also pointed out in~\cite{Prehofer2015}. FaaS-based functions have a single defined behaviour and can invoke the other functions by sending a request. These functions are stateless by design.  

The benefit of each modelling approach for IoT depends on the context. With the introduction of Web 2.0, consumers were invited to interact and engage in a community established around the contents that they consume~\cite{Ogrinz2009}. Now, these users are creating content themselves. Model-based and FaaS development are usually not the end-user's choice because they require a strong knowledge base and technical background. Mashup tools, on the contrary, make development accessible to a wider population. End-users who want to create IoT applications should not be forced to learn software development concepts. Instead, mashup tools designed for end-user should strive to continue opening the feature offering while keeping a low complexity of the user interface. Similarly, companies should allow their customers to extend and customize their applications, by exposing well documented APIs.

Enterprise applications demand fast but robust development that guarantees data privacy, security, and reliability. Model-based ideas are platform-independent, therefore the system design is not bounded to the system implementation. That removes constraints from the developer, who can later implement or generate the code in the most suitable language~\cite{Prehofer2015}. Additionally, model-based components support predictive maintenance, which ensures its optimum state. Mashups, on the other hand, are only suitable for critical applications if all the nodes are enterprise-managed resources. External resources might become unavailable unexpectedly, causing downtime and losses~\cite{Ogrinz2009}. In case of FaaS-based approach, the operational management is handled by the cloud service provider and hence the responsibility for providing a reliable service is their job. Though functions provide fault tolerance and high availability but there are some cases where a failure of function during a critical operation of an application can impact the performance. Furthermore, due to high virtualization stack in FaaS, they are no suitable for applications requiring nanoseconds or milliseconds of performance. 


Model-based techniques, have a range of reusability limited to code snippets, classes, and libraries. In mashups existing high-level structures such as databases, parsers, applications (through the API), etc., can be reused.  FaaS functions are inherently responsible for one task, therefore other functions requiring that particular task can trigger these functions leading towards the reusability of the functions. Each of the individual function can be scaled independently depending on the requirements. Furthermore, it offers three advantages  (i) no continuously running services are required, (ii) functions are only charged when they are executed, and (iii) the function abstraction increases the developer's productivity.

The need to manually define the behavior of a non-predefined component has been identified as a disadvantage of mashup development~\cite{Prehofer2015}. However, that might rather be an opportunity to integrate model-based with mashup and FaaS based approaches. Single components can be constructed using model-based or FaaS based development, and integrated using mashup tools. In that sense, the industry can greatly benefit from the combination of design principles in model-based techniques and the high-level reusability of components that characterizes mashups and FaaS based approaches. Furthermore, FaaS based approach also provides high scalability benefits which can be advantageous for certain applications.  


\section{Conclusion}
\label{sec_conclusion}
The Internet of Things (IoT) is a network where things can share information about their status and their surroundings, by establishing communication channels with other things and by enabling user interfaces. There is a large diversity of scenarios where  IoT technologies and protocols are implemented. As the coordination and communication between things is an essential factor, the development process of IoT applications is demanding. 

Model-based approaches offer structure and design principles that allow for an extensive description of the application from different abstraction levels. The design is not bounded to a specific platform because it is based on premises rather than technologies. Nevertheless, the process requires longer time-to-release given that the reuse of assets is limited, therefore, it enforces the implementation of every module. Code generation can be used to accelerate the process. The tools used for this purpose examine the system diagrams and produce matching lines of code.

Mashup development is faster than model-based development because it leverages existing resources at a high level. Different pieces of software can be integrated so that they can interact and deliver a new service or application. Mashups allow developers to focus on innovation, and non-programmer users to customize applications. A disadvantage of this development technique is that high availability, reliability, and performance can only be guaranteed when no resources come from external sources.

FaaS based development approach is similar to the mashup with the advantages of high availability, reliability, and scalability. Different components of an IoT application can be implemented as FaaS functions and each of the components can then be independently scaled. Due to high virtualization stack in a FaaS platform, this approach is not suitable for applications requiring nanoseconds or milliseconds of performance. 

A combination of these development approaches, in which the design of modules is accomplished using model-based principles, and their implementations are integrated using mashup tools or FaaS functions is an interesting alternative to be explored. Further work is required to evaluate the scalability and flexibility of this solution in the context of enterprise applications. 


\bibliographystyle{ACM-Reference-Format}
\bibliography{references}




\end{document}